\documentclass{aa}  

\usepackage{graphicx}
\usepackage{txfonts}
\usepackage{xcolor}
\begin{document}

    \title{A too-many dwarf satellite galaxies problem in the MATLAS low-to-moderate density fields}
    
    \author{Kosuke Jamie Kanehisa \inst{1,2}
        \and
        Marcel S. Pawlowski \inst{1}
        \and
        Nick Heesters \inst{3}
        \and
        Oliver Müller \inst{3}
        }
    
    \institute{Leibniz-Institut für Astrophysik Potsdam (AIP),
            An der Sternwarte 16, 14482 Potsdam\\
            \email{kkanehisa@aip.de}
        \and
            Institut für Physik und Astronomie, Universität Potsdam,
            Karl-Liebknecht-Straße 24/25, 14476 Potsdam, Germany
        \and
            Institute of Physics, Laboratory of Astrophysics, Ecole Polytechnique F\'ed\'erale de Lausanne (EPFL), 1290 Sauverny, Switzerland\\
            }
    
    \date{Accepted April 11, 2024}
 
  \abstract
   { 
   Dwarf galaxy abundances can serve as discernment tests for models of structure formation.
   Previous small-scale tensions between observations and dark matter-only cosmological simulations may have been resolved with the inclusion of baryonic processes; however, these successes have been largely concentrated on the Local Group dwarfs  the feedback models were initially calibrated on.}
   { 
   We investigate whether the $\Lambda$CDM model can reliably reproduce  dwarf abundances in the MATLAS low-to-moderate density fields that are centred upon early-type host galaxies beyond the Local Volume.}
   { 
   We carried out mock observations of MATLAS-like fields with the high-resolution hydrodynamic simulation IllustrisTNG-50. We used matching selection criteria and compared the properties of dwarfs contained within them with their MATLAS analogues.}
   { 
   Although simulated MATLAS-like dwarfs demonstrate photometric properties that are consistent with the observed galaxy population and follow the same scaling relations, TNG50 underestimates the number of dwarf galaxies in isolated MATLAS fields at the $6\sigma$ level. This significance is maintained within crowded fields containing more than a single bright host.
   Our $55-62\%$ estimate of the fraction of background galaxies is in agreement with estimates by MATLAS, but is wholly insufficient to alleviate this discrepancy in dwarf abundances.
   Any incompleteness in the observed fields further exacerbates this tension.
   }
   { 
   We identified a "too-many-satellites" problem in $\Lambda$CDM, emphasising the need for the continued testing and refining of current models of galaxy formation in environments beyond the Local Group.}


    \titlerunning{A too-many dwarf satellite galaxies problem in the MATLAS low-to-moderate density fields}
    \authorrunning{Kanehisa et al. 2024}

   \maketitle
%

\section{Introduction}
\label{sec:s1}

The concordance $\Lambda$CDM model has been largely successful in matching astrophysical observations at cosmological scales, such as the accelerated expansion of the Universe \citep{PlanckCollaboration2016}, the CMB power spectrum \citep{Spergel2007three-year}, and filamentary large-scale structure \citep{Tempel2014detecting}.
However, comparing the expectations from cosmological simulations with observed galaxies and their dwarf satellites has revealed a number of tensions at galactic scales \citep{Bullock2017small-scale, Sales2022baryonic}.

Abundances of dwarf galaxies are particularly sensitive to the feedback prescription adopted.
In the 'missing satellites' problem, dark matter-only simulations predict many more subhalos around Milky Way and M31-mass galaxies than what has been observed with respect to dwarf satellites \citep{Klypin1999where, Moore1999dark}.
This discrepancy arises from the quenching of galaxy formation in low-mass subhalos from gas heating due to the UV background from cosmic reionisation \citep{Bovill2009pre-reionization, Revaz2018pushing}.
Modern hydrodynamical and high-resolution $\Lambda$CDM simulations consistently match the luminosity function of the Milky Way and M31 dwarfs; thus, there no longer appears to be  a 'missing satellites' problem within the Local Group \citep{Engler2021abundance}.

Conversely, we are faced with a 'too-big-to-fail' (TBTF) problem \citep{Boylan-Kolchin2011too} with respect to the high-mass end of the subhalo mass function.
The dark matter masses of the brightest Milky Way satellites, as inferred from their internal velocities, are significantly lower than the most massive subhalos around simulated analogues \citep{Boylan-Kolchin2012milky}.
These dark subhalos are sufficiently massive to enable the effective cooling of their gas content and, thus, they are too big to fail in forming visible galaxies.
 Yet, with baryons, the increased host galaxy potential enhances the tidal stripping and mass loss of their satellites \citep{Brooks2014why}, while the formation of cored profiles effectively reduce the host halo mass and, accordingly, the expected subhalo abundance \citep{Peñarrubia2010impact}.
There is now a consensus among modern hydrodynamic simulations that there is no TBTF problem for Milky Way and M31-mass hosts \citep{Sawala2016apostle, Samuel2020profile}; however,  we refer to \citet{Pawlowski2015persistence} for an alternative viewpoint.

Since the baryonic solutions implemented in cosmological simulations are calibrated for Local Group-like environments, we highlight the need to verify whether these successes remain robust beyond our cosmic neighbourhood.
For instance, \citet{Smercina2018lonely} searched for dwarfs within $150\,\mathrm{kpc}$ of the Milky Way-mass spiral galaxy M94 ($D=4.2\,\mathrm{Mpc}$; \citealt{Radburn-Smith2011ghosts}) and found only two 'classical' satellites with $M_*>4\times10^5\,M_{\odot}$.
The four other comparably massive hosts with complete 'classical' samples host an average of 6-12 such satellites, and only $<0.2\%$ of M94 analogs in the EAGLE simulation \citep{Schaye2015eagle} host similarly sparse satellite systems.
Similarly, a study of seven low-mass spiral galaxies in the galactic neighbourhood \citep{Müller2020abundance} found that high-resolution dark matter simulations appear to systematically overestimate the observed satellite abundances (although this discrepancy may arise from observational biases).
Conversely, the satellite luminosity function around Centaurus A -- a massive elliptical galaxy at $3.8\,\mathrm{Mpc}$ that has been the target of multiple deep surveys -- is shown to be consistent with expectations from the hydrodynamic IllustrisTNG-100 simulation to within a $90\%$ confidence interval \citep{Müller2019dwarf}.

Recently, \citet{müller2024toomany} studied the abundance of dwarf galaxies around M83, a barred spiral at $D=4.9\,\mathrm{Mpc}$ \citep{Jacobs2009extragalactic}.
with a mass (and, thus, the expected satellite abundances) that is comparable to the Milky Way values.
M83 has 13 confirmed satellites within a projected separation of $330\,\mathrm{kpc}$ down to a limiting magnitude of $M_V=-10\,\mathrm{mag}$.
M83 analogs in the high-resolution IllustrisTNG-50 simulation, however, demonstrate sparse satellite distributions with luminosity functions inconsistent with that of the M83 system at a $3\sigma$ confidence level -- thus constituting the so-called 'too-many-satellites' problem.
The actual degree of tension is likely underestimated due to incompleteness in the observed satellite sample.

These ambiguous results suggest that models of galaxy formation in $\Lambda$CDM may still struggle to consistently predict the abundances of dwarf satellites around host galaxies beyond the Local Group. 
And yet, the works above only consider dwarf abundances of a few individual systems, which may conceivably be over- or under-populated as a result of some unique evolutionary history. 
In this paper, we study dwarf galaxy abundances within the low-to-moderate density fields in the MATLAS survey \citep{Duc2015atlas3d, Habas2020newly, Bilek2020census, Poulain2021structure}, across a total of 150 fields centred upon early-type host galaxies beyond the Local Volume.
We then mock-observe analogous fields using identical photometric selection criteria in the state-of-the-art hydrodynamic simulation IllustrisTNG-50, which features a resolution rivalling or even exceeding zoom-in simulations which consistently predict dwarf properties to the lower end of the subhalo mass function.
In this way, we tested for the first time whether dwarf abundances in MATLAS are consistent with expectations from concordance cosmology.

\section{Method}
\label{sec:s2}

\subsection{ATLAS$^{\mathrm{3D}}$ and the MATLAS survey}
\label{sec:s2_matlas_survey}

The Mass Assembly of early-Type gaLAxies with their fine Structures (MATLAS) deep imaging survey targets a volume-complete sample of early-type galaxies (ETGs) acquired from the ATLAS$^{\mathrm{3D}}$ legacy program \citep{Cappellari2011atlas3d}.
ATLAS$^{\mathrm{3D}}$ consists of 260 elliptical and lenticular galaxies with K-band absolute magnitudes of $M_K < -21.5$, selected from a parent sample of 871 massive galaxies.
The targeted ETGs are located at declinations of $|\delta - 29^{\circ}| < 35^{\circ}$ and Galactic latitudes of $b > 15^{\circ}$, and lie within $10-45\,\mathrm{Mpc}$ ($z<0.01$) of the Milky Way.
Together with the Next Generation Virgo cluster Survey \cite[NGVS;][]{Ferrarese2012next}, MATLAS conducted deep optical imaging of the regions around the ATLAS$^{\mathrm{3D}}$ ETGs.
NGVS mapped fields around 58 of the ATLAS$^{\mathrm{3D}}$ targets between 2009 and 2013, while MATLAS has since imaged most of the remaining ETGs.

The MATLAS survey consists of a total of 150 fields around 180 ETGs and 59 late-type galaxies (LTGs) from ATLAS$^{\mathrm{3D}}$'s parent catalogue, each with dimensions of $63'\times69'$ and a resolution of $0.187\,\mathrm{arcsec}\,\mathrm{pix}^{-1}$.
Multiband observations were performed using the MegaCam on the 3.6-metre Canada-France-Hawaii Telescope (CFHT) between 2010 and 2015 \citep{Duc2015atlas3d}.
MATLAS was designed to study low surface brightness features in the outskirts of the ATLAS$^{\mathrm{3D}}$ ETGs to a local depth in the g-band of $\mu_g=28.5-29\,\mathrm{mag}\,\mathrm{arcsec}^{-2}$ \citep{Duc2014identification}; therefore, it is ideal for resolving populations of faint and low-surface brightness dwarf galaxies around their massive hosts.

\subsection{ MATLAS dwarf catalogue}
\label{sec:s2_matlas_dwarfs}

MATLAS adopts both a fully visual and semi-automatic approach to generate its dwarf catalogue.
The visual approach is used purely to calibrate the dwarf selection criteria used by the latter and the authors identified 1349 dwarfs across all 150 fields.
The semi-automatic catalogue first uses \texttt{SOURCE EXTRACTOR} (hereafter \texttt{SEXTRACTOR}; \citealt{Bertin1996sextractor}) to generate an automated list of dwarf candidates, followed up by a multi-stage visual cleaning procedure.
It was found that a linear cut in average surface brightness, $\langle \mu_g \rangle,$ and apparent magnitude, $m_g$, resulted in the cleanest separation between the visually selected dwarfs and non-dwarf detections from a representative field around NGC 4382 \citep{Habas2020newly}.
Substructures could only be identified in more massive galaxies -- hence confirming their background nature -- when their area was larger than $75\,\mathrm{pix}^2$ ($23.5\,\mathrm{arcsec}^2$) and all smaller objects were removed from the selection.
The catalogue was then cleaned manually by the authors and dwarfs were classified by their visual morphology.

The final dwarf catalogue contains 2210 unique dwarf candidates and is dominated by dwarf ellipticals (dE, $73.4\%$), while the remaining $26.6\%$ are classified as dwarf irregulars (dIrr).
One of the 150 MATLAS fields is devoid of detectable dwarfs due to intense light contamination from a nearby star.
The remaining 149 fields host between 2 and 79 dwarf candidates per field. Furthermore,
67 fields contain no bright galaxies (defined by a magnitude of $<M_K+1$), other than the field's target ETG, while a further 22 fields lack bright galaxies in close projection to the central ETG within a square 175 kpc region.

Dwarf galaxies are generally assumed to be satellites of the central ETGs in their respective fields.
There are 325 dwarf candidates ($15\%$) with at least one independent distance estimate, spanning a range between $5.1$ and $100.6\,\mathrm{Mpc}$. Furthermore,
$90\%$ of these dwarfs lie within the $10-45\,\mathrm{Mpc}$ region populated by the ATLAS$^{\mathrm{3D}}$ ETGs.
Intrinsic properties including effective radius, $R_e$, Sersic index, $n$, and axis ratio, $b/a,$ were derived using \texttt{GALFIT} \citep{Peng2010detailed} in \citet{Poulain2021structure}.
Since a majority of the MATLAS dwarfs lack associated distances, their properties were calculated under the assumption that each dwarf is located at the same distance as their field's targeted host galaxy.
The authors confirm the dwarf nature of the catalogued objects using a cut of $M_g > -18$, and $80\%$ of dwarfs appear to be physically associated to their field's central ETG with a relative radial velocity within $500\,\mathrm{km}\,\mathrm{s}^{-1}$.
Of 56 dwarfs with additional MUSE spectroscopy, 42 ($75\%$) demonstrate radial velocities within $1000-3000\,\mathrm{km}\,\mathrm{s}^{-1}$, thus consistent with the ATLAS$^{\mathrm{3D}}$ galaxies \citep{Heesters2023radial}.

\begin{figure*}
    \centering
    \includegraphics[width=\textwidth]{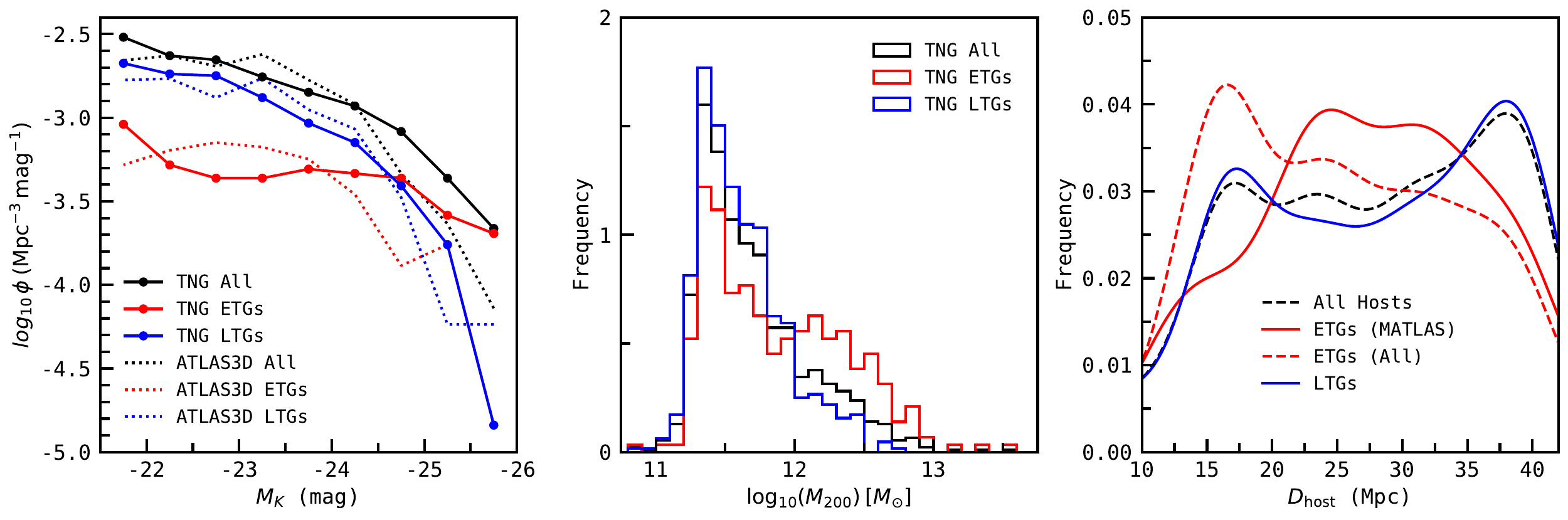}
    \caption{
        Properties of ATLAS$^{\mathrm{3D}}$ host galaxies and their TNG50 analogues. \textbf{Left:} Luminosity functions of TNG50 and ATLAS$^{\mathrm{3D}}$ hosts are plotted in solid and dotted lines, respectively. The latter corresponds to Fig. 3 in \citet{Cappellari2011atlas3d}. Samples containing only early-type and late-type centrals are denoted in red and blue respectively. \textbf{Middle:} Mass distribution of the dark matter halos hosting the TNG50 analogues. \textbf{Right:} Distance distribution of ATLAS$^{\mathrm{3D}}$ targets, smoothened via a Gaussian kernel for re-sampling in this work. The MATLAS survey targets a subset of the early-type ATLAS$^{\mathrm{3D}}$ hosts (solid red line).
    }
    \label{fig:host_properties}
\end{figure*}

\subsection{Host galaxy selection in TNG50}
\label{sec:s2_tng_hosts}

We use data from the IllustrisTNG suite\footnote{Publicly available at \texttt{www.tng-project.org}.} of large-volume hydrodynamic cosmological simulations \citep{Pillepich2018simulating}.
All runs contained therein adopt cosmological parameters taken from Planck \citep{PlanckCollaboration2016}: $\Omega_{\Lambda} = 0.6911$, $\Omega_\mathrm{m}=0.3089$, and $h=0.6774$.
We specifically used the highest-resolution TNG50 run ($L_{\mathrm{box}}=51.7\,\mathrm{Mpc}$) which (with a baryonic particle mass of $m_{\mathrm{gas}}=8.5\times10^4M_{\odot}$)  best resolves populations of MATLAS-like dwarfs with estimated stellar masses of $5.8<\mathrm{log}_{10}(M_*/M_{\odot})<9$ \citep[see][]{Habas2020newly}.

IllustrisTNG adopts the classic friends-of-friends (FOF) approach to identify virialised dark halos, then populating them with subhalos using \texttt{SUBFIND} \citep{Springel2001populating}.
To select ATLAS$^{\mathrm{3D}}$-like host galaxies in TNG50, we first identified all subhalos considered by \texttt{SUBFIND} to be the central subhalo of some parent halo.
ATLAS$^{\mathrm{3D}}$-like hosts are selected by mirroring \citet{Cappellari2011atlas3d}'s criterion and adopting a K-band absolute magnitude cut of $M_K<-21.5$.
Since TNG50 contains several highly luminous galaxies up to $M_K \sim -28$ that lack analogues in the parent ATLAS$^{\mathrm{3D}}$ catalogue, we imposed an additional constraint of $M_K>-26.0$ and recover a total of 925 TNG hosts.

We classified host galaxies by their Hubble morphology into early-types (ETGs) and late-types (LTGs).
This distinction is made by referring to \citet{Zana2022morphological}'s kinematic decomposition of $M_*>10^9M_{\odot}$ subhalos in TNG50 into five distinct components -- the thin disc, thick disc, pseudo-bulge, bulge, and stellar halo.
While the original work defines a given galaxy's disc-to-total mass ratio $D/T$ as the combined mass fraction of the former three rotating components, we argue that only the thin and thick disc would be observationally distinguishable; thus, we excluded the pseudo-bulge from our $D/T$ calculation.
Since there is no clear distinction between ETGs and LTGs in this metric, we took advantage of ATLAS$^{\mathrm{3D}}$'s volume-complete nature and assume the fraction of observed ETGs ($260/871=0.299$) also holds in TNG50.
The 30th percentile of the full $D/T$ distribution at $0.575$ is hence set as the threshold between early-types and late-types, and we identify 287 ATLAS$^{\mathrm{3D}}$-like ETGs.

The properties of the sampled ATLAS$^{\mathrm{3D}}$-like hosts in TNG50 are shown in Fig.~\ref{fig:host_properties}. The simulated luminosity function roughly agrees with that in ATLAS$^{\mathrm{3D}}$ \citep{Cappellari2011atlas3d}, although TNG50 demonstrates an excess of luminous early-type hosts at $M_K < -24$. The sample of TNG50 analogues appears to be dominated by Milky-Way and M31-mass objects within halos of $10^{11.5}\,M_{\odot} < M_{200} < 10^{12.5}\,M_{\odot}$.

\subsection{Mock-observing MATLAS-like fields}
\label{sec:s2_mock_observing}

We mock-observed $63'\times69'$ fields centred upon the selected TNG hosts to resemble the MATLAS low-to-moderate density fields as closely as possible.
ATLAS$^{\mathrm{3D}}$'s distribution of the 180 ETGs targeted by MATLAS \citep{Cappellari2011atlas3d} were smoothed using a Gaussian kernel. The obtained kernel density estimate (KDE) was truncated at $10\,\mathrm{Mpc}$ and $45\,\mathrm{Mpc}$ to mirror the observed sample (see the right-hand panel in Fig.~\ref{fig:host_properties}).
We generate $N_{\mathrm{real}}=10$ realisations for each early-type TNG host galaxy, each with a mock-observation distance, $D_{\mathrm{host}}$, independently drawn from the ATLAS$^{\mathrm{3D}}$ KDE.

For each realisation, we drew an observing direction, $\hat{r,}$ from an isotropic distribution and define an associated cone with an opening angle of $\theta_{\mathrm{field}}=46.7\,\mathrm{arcmin}$ (thus fully containing the final $63'\times69'$ survey footprint centred upon the target host) and observation depth, $D_{\mathrm{obs}}$, aligned with $\hat{r}$.
An observer is placed at a distance of $D_{\mathrm{host}}$ from each ATLAS$^{\mathrm{3D}}$-like host galaxy along $-\hat{r}$ and we proceeded to mock-observe all luminous subhalos within a $63'\times69'$ field around the host across multiple periodic volumes until $D_{\mathrm{obs}}=100\,\mathrm{Mpc}$ was reached.

Since it is unfeasible to straightforwardly replicate \citet{Habas2020newly}'s visual inspection of dwarf candidates for the mock-observed subhalos, we focussed on adopting the quantitative selection criteria used to compile the final MATLAS dwarf sample.
The greatest challenge lies in estimating effective (half-light) radii $R_e$ for TNG's luminous subhalos, a rigorous approach to which would require the individual modelling of Sérsic profiles for each subhalo while accounting for their relative orientation and morphology.
We instead approximated $R_e$ with a straightforward conversion from stellar half-mass radii $R_{1/2}$, which is provided by TNG's \texttt{Subfind} group catalogues.
\citet{Price2022kinematics} found that the ratio between $R_{1/2}$ and $R_e$ is generally consistent over varying Sérsic indices and, rather, it primarily depends on the galaxy's projected axis ratio, $q_0$.
The full population of MATLAS dwarfs \citep{Habas2020newly} demonstrates a distribution of $q_0$ between 0.3 and 1.0, with a median value of roughly $q_0=0.7$.
The range of MATLAS $q_0$ from \texttt{GALFIT} covers $1.05<R_{1/2}/R_e<1.3$, and we adopt a corresponding static $R_{1/2}/R_e$ ratio of 1.2 (see Fig.~2 in \citealt{Price2022kinematics}).

The apparent g-band magnitude of TNG's luminous subhalos is derived from their absolute magnitude $M_g$ in the usual manner using:
\begin{equation}
    m_g = M_g + 5\mathrm{log}_{10}(D/10\mathrm{pc})
    \label{eq:app_mag}
,\end{equation} 
when mock-observed at a distance $D$, although we do not account for the effects of dust attenuation.
Each subhalo's average surface brightness $\langle \mu_g \rangle$ within $R_e$ (in units of $\mathrm{mag}\,\mathrm{arcsec}^{-2}$) is calculated from $m_g$ and the angular expression of $R_e$ as:
\begin{equation}
    \langle \mu_g \rangle = 0.75 + m_g + 2.5\mathrm{log}_{10}(\pi\,(R_e/\mathrm{arcsec})^2),
    \label{eq:sb}
\end{equation}
where the factor $2.5\log_{10}(2)=0.75$ arises from considering the flux observed only within $R_e$.

\begin{figure*}
    \centering
    \includegraphics[width=\textwidth]{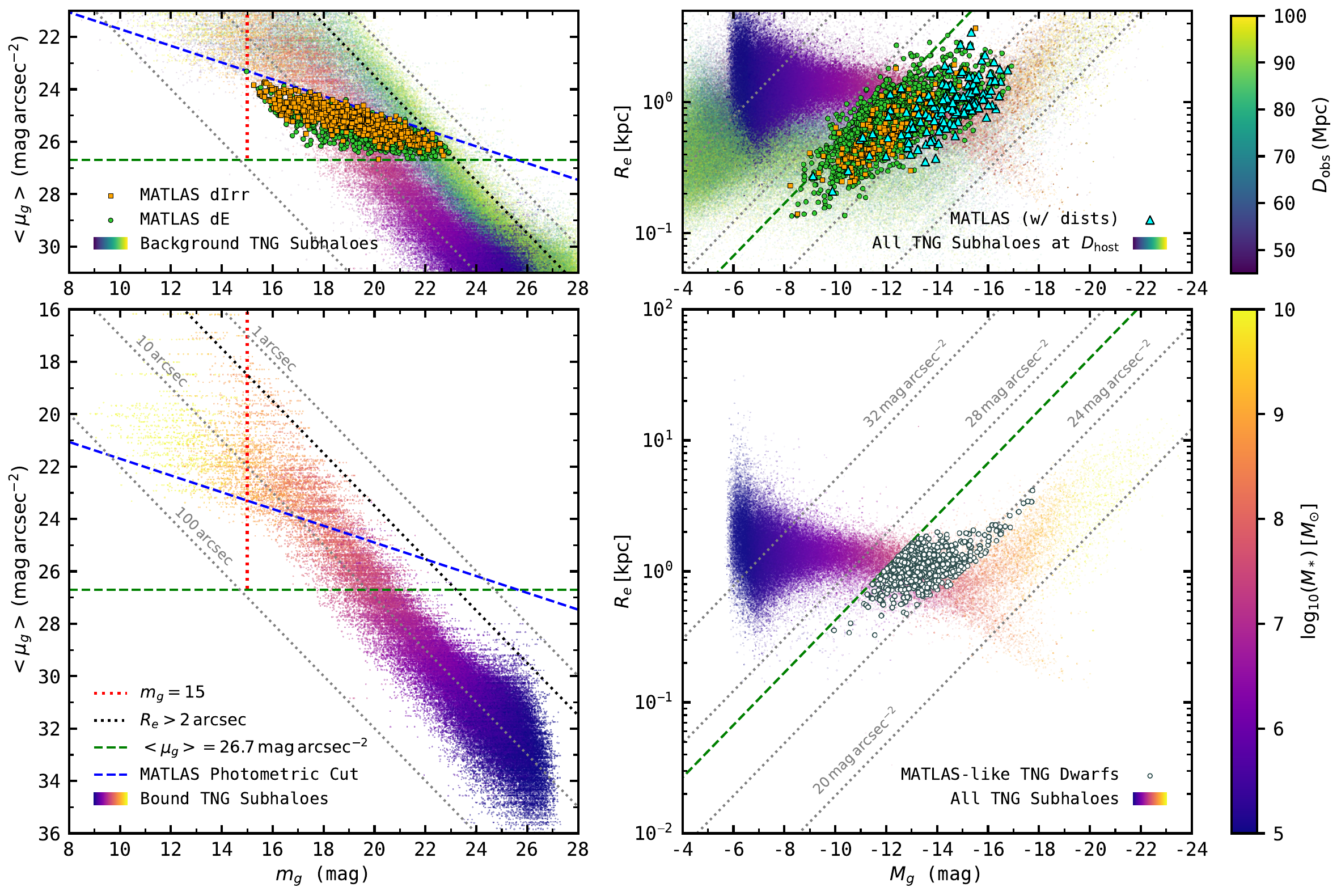}
    \caption{
    Scaling relations for MATLAS dwarfs and luminous subhaloes in the TNG50 simulation.
    \textbf{Left:} Selection criteria for MATLAS-like dwarfs mock-observed in TNG50.
    The blue line indicates the linear cut between regular galaxies and dwarfs adopted in MATLAS \citep{Habas2020newly}.
    The green, red, and black lines respectively represent cuts on average surface brightness $\langle \mu_g \rangle$, apparent magnitude $m_g$, and apparent effective radius motivated by the distribution of the MATLAS dwarfs (see upper panel).
    The observed dwarfs demonstrate photometric properties consistent with their TNG counterparts, except for a population at larger $m_g$ than expected in the simulated $\langle \mu_g \rangle - m_g$ relation.
    This discrepancy can be explained by assuming that a fraction of the MATLAS dwarfs are background galaxies with distances up to $100\,\mathrm{Mpc}$.
    \textbf{Right:} Physical properties of luminous subhalos in TNG's mock-observed fields.
    MATLAS-like dwarfs that satisfy the selection criteria are indicated as white circles.
    Distances are not available for around $85\%$ of the MATLAS dwarfs, and they are assumed to lie at the same distance as their field's target host galaxy.
    The TNG dwarfs, when making this assumption, also follow a similar extended distribution in $R_e - M_g$ space.
    MATLAS dwarfs with known distances follow the true TNG distribution more closely.
    }
    \label{fig:dwarf_properties}
\end{figure*}

\subsection{Selection criteria for MATLAS-like dwarfs}
\label{sec:s2_selection_criteria}

The distribution of observed MATLAS dwarfs in $\langle \mu_g \rangle - m_g$ space is plotted in the upper left panel of Fig.~\ref{fig:dwarf_properties}.
We differentiate between dwarfs and more massive galaxies by implementing \citet{Habas2020newly}'s linear photometric cut of
\begin{equation}
    \langle \mu_g \rangle \geq 0.32 m_g + 18.5.
    \label{eq:linear_cut}
\end{equation} 
While the authors calibrate this cut for each field to best separate the low-surface-brightness region of the main locus of galaxies with the visually confirmed MATLAS dwarfs, these shifts generally remain within $\langle \mu_g \rangle$ offsets of less than $0.1\,\mathrm{mag}\,\mathrm{arcsec}^{-2}$.
Hence, we adopted Eq.~\ref{eq:linear_cut} (blue dashed line) as a static threshold for the remainder of this work.
While MATLAS is capable of detecting substructures to local depths of $\mu_g = 28.5-29\,\mathrm{mag}\,\mathrm{arcsec}^{-2}$, the lowest average surface brightness within $R_e$ recorded within the final dwarf catalogue is only $26.7\,\mathrm{mag}\,\mathrm{arcsec}^{-2}$.
Hence, we defined a criterion for average surface brightness of: 
\begin{equation}
    \langle\mu_g\rangle<26.7\,\mathrm{mag}\,\mathrm{arcsec}^{-2}
    \label{eq:sb_cut}
,\end{equation}
for dwarfs in the mock-observed TNG fields (green dashed line).

The region in $\langle \mu_g \rangle - m_g$ space encompassed by the two cuts defined above contains a population of visually bright galaxies at $m_g<15$ that is not present in the MATLAS catalogue.
Their absence is likely due to the visual inspection by \citet{Habas2020newly} correctly identifying them as non-dwarfs.
Since all MATLAS dwarfs demonstrate \texttt{GALFIT} and \texttt{SEXTRACTOR} magnitudes of $m_g>14.86$, we corrected for this discrepancy by imposing an additional, conservative cut at:
\begin{equation}
    m_g>15,
    \label{eq:mag_cut}
\end{equation} 
(red dotted line). Finally, MATLAS rejects all dwarf candidates with an \texttt{ISOAREA} parameter in \texttt{SEXTRACTOR} of $75\,\mathrm{pix}^{2}$.
While \texttt{ISOAREA} has no exact physical analogue available for TNG subhalos, we find that this rejection cut corresponds to a minimum angular effective radius of:
\begin{equation}
    R_e>2\,\mathrm{arcsec,}
    \label{eq:reff_cut}
\end{equation} 
(black dotted line) in the MATLAS catalogue, which we adopted as our last selection criteria for MATLAS-like dwarfs. This cut in $R_e$ serves to further remove background interlopers from our sample, as only $<0.1\%$ of physically bound MATLAS-like dwarfs are rejected in this final step.

The population of galaxies in our mock-observed TNG fields consist of satellite galaxies of the ETGs targeted by each field, as well as a fraction of interlopers -- foreground or background objects that are not physically associated with their field's central galaxy.
An accurate distinction can be made by whether a MATLAS-like dwarf is considered to be bound to their presumed host galaxy's halo in \texttt{SUBFIND}'s halo catalogue.
A more observationally analogous approach uses the difference in radial velocities between the given dwarf and its field's central ETG while taking their recession velocities due to the Hubble flow into account.
\citet{Habas2020newly} estimated that using a cut of $|\Delta v| < 500\,\mathrm{km}\,\mathrm{s}^{-1}$, around $80\%$ of dwarf candidates with independent distance estimates appear to be satellites of the ETG or LTG with the smallest angular separation.
We adopted the same threshold below to test whether we obtain a comparable fraction of physically associated dwarfs in TNG's MATLAS-like fields.

\begin{figure*}
    \centering
    \includegraphics[width=\textwidth]{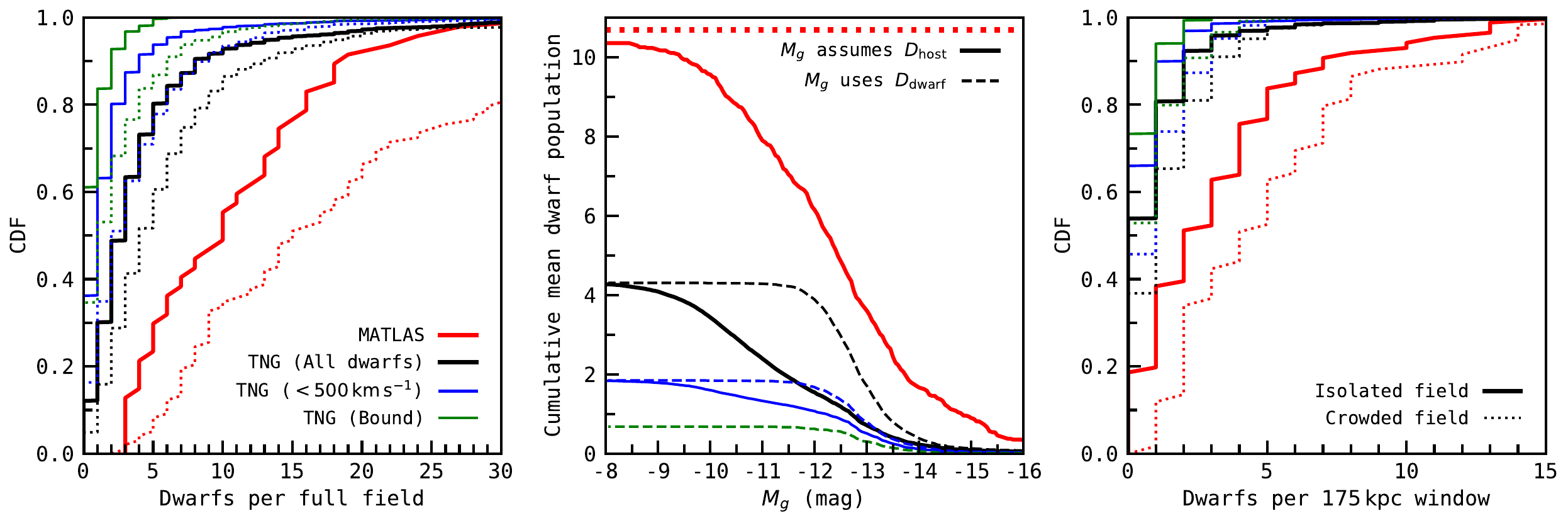}
    \caption{
    Population of MATLAS-like dwarfs in TNG's mock-observed fields is significantly lower than observed in the MATLAS low density fields.
    \textbf{Left:} Dwarf counts in the $63'$ by $69'$ MATLAS fields (red) and MATLAS-like fields mock-observed in TNG50 (black).
    Dwarf galaxies that appear to be physically associated with the targeted hosts (by a radial velocity difference of $<500\,\mathrm{km}\,\mathrm{s}^{-1}$) are also shown in blue, while dwarfs bound to their host galaxy's halo are drawn in green.
    Solid lines indicate fields around isolated MATLAS and MATLAS-like host galaxies that lack a companion with a magnitude of $< M_K + 1$ within $1\,\mathrm{Mpc}$, while dotted lines represent the remaining fields.
    Dwarf populations of isolated TNG fields are inconsistent with these isolated MATLAS fields -- even when including unassociated interlopers in the former -- at a $7\sigma$ confidence level.
    \textbf{Centre:} Cumulative luminosity functions for MATLAS and TNG dwarfs in isolated fields.
    The solid line represents dwarf absolute magnitudes derived by assuming they lie at their supposed host galaxy's distance (like the MATLAS dwarfs), while the dashed line adopts their true distances instead.
    Since around $3\%$ of the MATLAS dwarfs lack $m_g$ estimates, the red dotted line indicates the true mean population of MATLAS dwarfs per field.
    \textbf{Right:} Same as the left panel, but only counting dwarfs that lie within a square field with physical dimensions of $175\,\mathrm{kpc}$ centred upon each targeted host galaxy corresponding to the MATLAS field size at the near bound of $10\,\mathrm{Mpc}$.
    }
    \label{fig:dwarf_population}
\end{figure*}

\section{Results \& discussion}
\label{sec:s3}

\subsection{Properties of MATLAS-like dwarfs}
\label{sec:s3_properties}

Mock-observed properties of TNG50 subhalos are plotted in the left-hand panels of Fig.~\ref{fig:dwarf_properties}.
The distribution of all luminous TNG subhalos bound to their field's central host galaxy follows a roughly linear relation in $\langle\mu_g\rangle-m_g$.
Dwarfs considered to be MATLAS-like lie in the region enclosed by the four selection criteria in Eqs.~\ref{eq:linear_cut}-\ref{eq:reff_cut} and exhibit stellar masses of the order of $10^7-10^8\,M_{\odot}$.
The observed MATLAS dwarfs in the upper panel inhabit a region consistent with our mock-observed analogues, but extend to higher $m_g$ for a given $\langle\mu_g\rangle$.
This shift can be accounted for if a fraction of the MATLAS dwarfs are background interlopers unassociated with their supposed host galaxies.
In Fig.~\ref{fig:dwarf_properties}, we demonstrate that these outlying MATLAS dwarfs are consistent with a background distribution at distances up to $90-100\,\mathrm{Mpc}$.
This maximum is in excellent agreement with the $13\%$ of MATLAS dwarfs with independent distance estimates -- which cover a range of distances up to $100.6\,\mathrm{Mpc}$ -- and justifies our choice of observation depth of $D_{\mathrm{obs}}=100\,\mathrm{Mpc}$ in Section~\ref{sec:s2_mock_observing}.

The right-hand panels of Fig.~\ref{fig:dwarf_properties} shows intrinsic properties of all luminous TNG subhalos.
MATLAS-like dwarfs (plotted as white circles) appear to follow the standard scaling relation between $M_g$ and $R_e$.
The full distribution of observed MATLAS dwarfs partially lies along the region inhabited by TNG's MATLAS-like analogues, but also appreciably extends to larger $M_g$ and lower $R_e$.
Since $85\%$ of MATLAS dwarfs lack distance estimates, both $R_e$ and $M_g$ were calculated by assuming they are located at the same distance as their supposed host galaxy \citep{Habas2020newly}.
When making this assumption for the luminous TNG subhalos, they span a much larger region in $M_g-R_e$ space that is now consistent with the MATLAS dwarfs.
The fact that the MATLAS dwarfs also extend beyond the $\langle\mu_g\rangle=26.7\,\mathrm{mag}\,\mathrm{arcsec}^{-2}$ threshold (despite following this cut in the left-hand panels) also hints at the unreliability of their size and absolute magnitude estimates.
The few MATLAS dwarfs with distance estimates (blue triangles) demonstrate a much closer alignment with TNG's MATLAS-like dwarfs and generally do not extend beyond the $\langle \mu_g \rangle$ threshold.
Overall, the MATLAS-like dwarfs mock-observed in TNG50 display photometric and intrinsic properties that are fully consistent with dwarfs observed in the MATLAS fields, especially when assuming a degree of contamination by background interlopers.

\subsection{MATLAS-like dwarf populations}
\label{sec:s3_populations}

The number of dwarf galaxies contained within the MATLAS fields depends heavily on the central host galaxy's degree of isolation.
Crowded fields containing multiple ETGs and/or LTGs have enhanced dwarf counts due to the stacking of multiple distinct satellite populations.
To control for this bias, we focus on fully isolated fields wherein the central ETG has no other bright galaxy with a magnitude of $<M_K+1$ within a 3D distance of $1\,\mathrm{Mpc}$ -- a threshold selected to distinguish isolated hosts from the Local Group pair $\sim 800\,\mathrm{kpc}$ apart, see Appendix~\ref{app:isolation}.
We disregarded galaxies without apparent k-band magnitudes since they are likely to lie on the fainter, low-mass end of the luminosity function.
We searched for potential companion galaxies in the Heraklion Extragalactic Catalogue (HECATE; \citealt{Kovlakas2021heraklion}), a value-added all-sky galaxy catalogue containing over 200,000 galaxies within 200 Mpc. Thus, we identified 48 fully isolated MATLAS fields that satisfy these criteria.

The left-hand panel in Fig.~\ref{fig:dwarf_population} plots the cumulative distribution of dwarf populations in the isolated MATLAS fields (solid red line).
We identified isolated MATLAS-like fields in TNG50 using the same criteria and plot their dwarf counts in black.
In stark contrast to the MATLAS median of ten dwarfs per isolated field, the mock-observed TNG fields only contain a median population of three MATLAS-like dwarf galaxies.
A two-sample Kolmogorov-Smirnov (KS) test reveals that the two distributions are inconsistent with being drawn from the same parent distribution at a $6.7\sigma$ confidence level ($D=0.503$, $p=2.9\times10^{-11}$).
This discrepancy only widens when solely considering TNG dwarfs with radial velocities within $500\,\mathrm{km}\,\mathrm{s}^{-1}$ of their field's central ETG; thus, MATLAS-like fields in TNG50 are found to only contain a median of one such dwarf.
When restricting our search to dwarfs bound to their field's host galaxy, the median satellite population drops to zero.
While we focus on isolated fields for this analysis, it is evident (see Fig. 3) that more crowded fields in TNG and MATLAS demonstrate a similar or greater degree of tension.
We also note that this result is robust to our adopted threshold between early-type and late-type hosts, as the latter ($D/T > 0.575$) demonstrate median populations of MATLAS-like dwarfs identical to the former.

\begin{figure*}
    \centering
    \includegraphics[width=\textwidth]{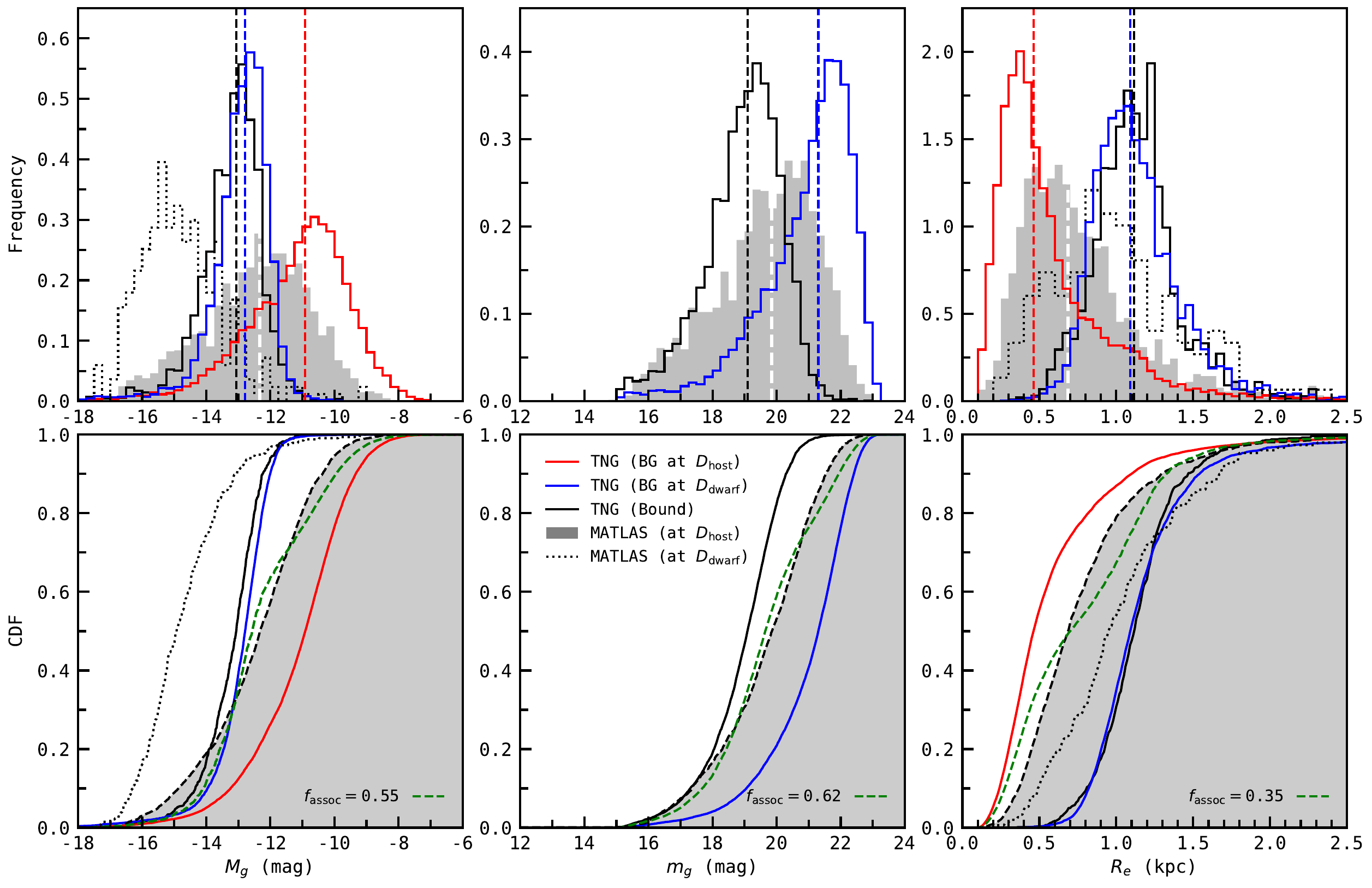}
    \caption{
    Properties of the MATLAS dwarfs and their simulated TNG analogues.
    TNG dwarfs bound to their host galaxy are plotted in black, while those unassociated with their supposed host are drawn in blue.
    MATLAS dwarfs are shown in shaded grey.
    Absolute magnitude, $M_g$, and effective radius, $R_e$, require distance estimates to derive, while apparent magnitude, $m_g$, is a purely observational property.
    For the former two, estimates derived by assuming TNG dwarfs lie at their designated host galaxy's distance are plotted in red.
    Of the MATLAS dwarfs, $15\%$ have individual distance estimates. The
    $M_g$ and $R_e$  values calculated using these known distances are drawn as black dotted lines.
    The distribution of MATLAS dwarfs in $M_g$, $m_g$, and $R_e$ generally lies between the bound and background population in TNG.
    We also estimate the fraction of MATLAS dwarfs physically associated to their field's host galaxy, $f_{\mathrm{assoc}}$, by averaging over TNG's bound and background CDFs (green dashed line) with a ratio that minimizes the corresponding Kolmogorov-Smirnov statistic.
    }
    \label{fig:dwarf_background}
\end{figure*}

Dwarf galaxy populations in the MATLAS fields can also be described using luminosity functions (Fig.~\ref{fig:dwarf_population}, central panel).
Due to the large variance in the number of dwarfs per field and the discrete nature of dwarf counts, we characterised the isolated MATLAS luminosity function (red line) using the cumulative mean dwarf population brighter than a given $M_g$. We note that
$2.8\%$ of MATLAS dwarfs lack $m_g$ estimates from \texttt{SEXTRACTOR} or \texttt{GALFIT}, so we indicate the true mean dwarf population per isolated field with a red dotted line.
As discussed in Section~\ref{sec:s3_properties}, $M_g$ is calculated for MATLAS dwarfs by assuming they lie at their presumed host galaxy's distance, $D_{\mathrm{host}}$.
Solid black and blue lines correspond to TNG dwarfs under the same assumption, while dashed lines show their true $M_g$ distribution.
Solid and dashed lines are roughly equivalent for bound TNG dwarfs (green line) due to their physical proximity to their field's central ETG and, thus, only one line is shown.
To fully match TNG's mean value of 0.74 bound satellites per isolated field, $93\%$ of the MATLAS dwarfs would need to be interlopers unassociated with their supposed host.

The one-degree fields in MATLAS cover a region around their central host galaxy that scales linearly with observation distance -- with a length ranging from $175\,\mathrm{kpc}$ at $D_{\mathrm{host}}=10\,\mathrm{Mpc}$ to nearly $800\,\mathrm{kpc}$ at the maximum host distance of $45\,\mathrm{Mpc}$.
To compensate for this dependence, we follow \citet{Habas2020newly}'s method and define a square window with dimensions of $175\,\mathrm{kpc}$, corresponding to the minimum physical field size around each MATLAS ETG.
The distribution of dwarfs located within this $175\,\mathrm{kpc}$ window is plotted in the right-hand panel of Fig.~\ref{fig:dwarf_population}.
In contrast to the isolated MATLAS median of two dwarfs per window, over $50\%$ of isolated MATLAS-like fields in TNG50 contain no dwarfs within their $175\,\mathrm{kpc}$ window.
According to the 2-sample KS test, this discrepancy constitutes a $7.3\sigma$ tension between the two distributions ($D=0.398$, $p=3.1\times10^{-13}$), a significance even exceeding the tension over the full one-degree fields.

\subsection{Satellite galaxies and interloper contamination}
\label{sec:s3_background}

When comparing populations of dwarf galaxies in the MATLAS fields and TNG's mock-observed analogues, the fraction of interloping background galaxies cannot be assumed to be similar a priori.
The multi-stage visual inspection process in the MATLAS survey \citep{Habas2020newly} can rule out a majority of highly luminous non-dwarfs due to their distinct morphology and the presence of substructure.
In TNG, however, we are limited to the use of MATLAS-like photometric cuts (Eqs.~\ref{eq:linear_cut}-\ref{eq:reff_cut}) to attempt to achieve the same result.

\begin{figure*}
    \centering
    \includegraphics[width=0.409\textwidth]{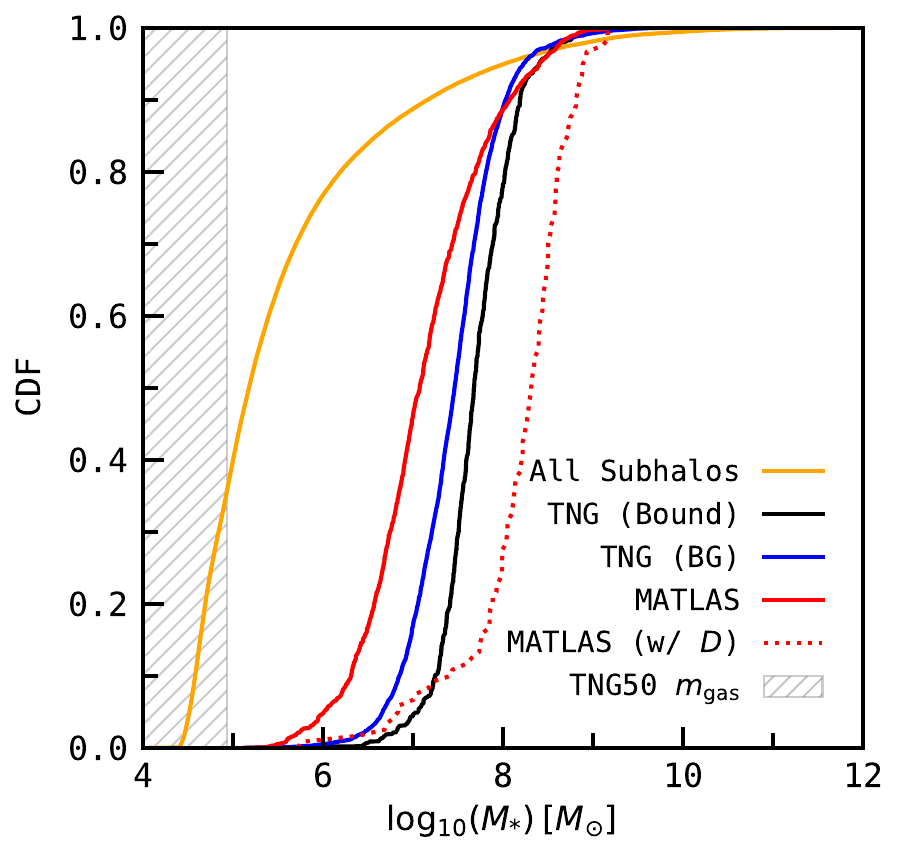}
    \hspace{5mm}
    \includegraphics[width=0.472\textwidth]{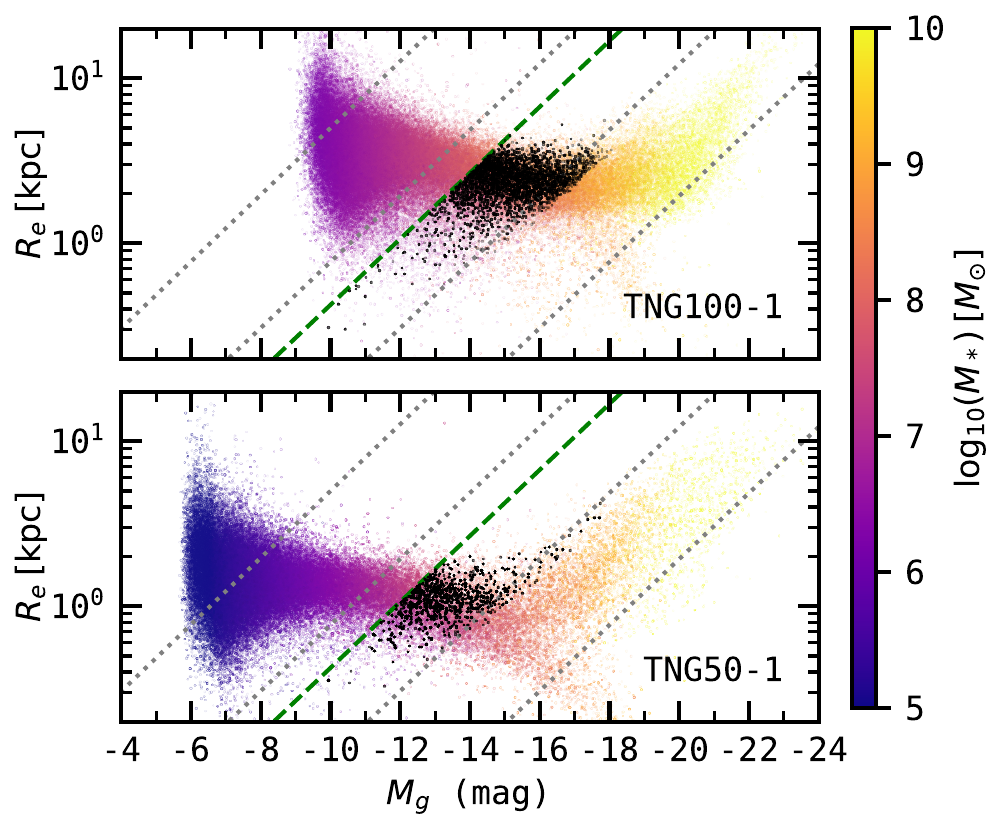}
    \hspace{3mm}
    \caption{
    \textbf{Left:} Stellar masses of TNG subhalos by their association to their supposed host galaxy (black and blue lines) and estimates for the MATLAS dwarfs from \citep[red line;][]{Habas2020newly}.
    The latter estimates were made by assuming that the observed dwarfs lie at the same distance as their field's target host galaxy.
    Stellar masses for the subset of MATLAS dwarfs for which independent distance estimates are available are plotted as a red dotted line.
    TNG50's baryonic resolution $m_{\mathrm{gas}}$ is indicated by the hatched region.
    $97\%$ of MATLAS-like dwarfs in TNG50 are sufficiently resolved with at least 100 stellar particles.
    \textbf{Right:}  Impact of simulation resolution on dwarf populations.
    The lower panel is taken from Fig.~\ref{fig:dwarf_properties} and shows the distribution of luminous subhalo properties in $R_e - M_g$ space in TNG50 ($m_{\mathrm{gas}}=8.5\times10^4\,M_{\odot}$), while the upper panel instead displays results for the lower-resolution TNG100 run ($m_{\mathrm{gas}}=1.4\times10^6\,M_{\odot}$).
    MATLAS-like dwarfs satisfying the selection criteria are plotted in black.
    The broadening in $R_e$ due to insufficient stellar particles occurs at around $M_g > -12$ in TNG100, but only affects $M_g > -9$ in TNG50 at a regime much fainter than the MATLAS-like dwarfs.
    }
    \label{fig:dwarf_resolution}
\end{figure*}

\citet{Habas2020newly} reported that $77-82\%$ of MATLAS dwarfs with H1 velocities or distance estimates demonstrate radial velocities within $500\,\mathrm{km}\,\mathrm{s}^{-1}$ of their fields' central host galaxies.
On the other hand, \citet{Heesters2023radial} recently obtained additional spectroscopy for 56 MATLAS dwarfs from VLT's Multi Unit Spectroscopic Explorer (MUSE; \citealt{Bacon2010muse}),  finding that a much lower estimate of $57\%$ is likely associated with their respective fields' central ETGs using $|\Delta_v| < 500\,\mathrm{km}\,\mathrm{s}^{-1}$.
In TNG's mock-observed fields, a similar satellite fraction of $55\%$ of all MATLAS-like dwarfs ($43\%$ in isolated fields) satisfies the same $|\Delta_v|$ criterion, although the true fraction of dwarfs bound to their field's central ETG is much lower at $31\%$ ($16\%$).
This counter-intuitive result of isolated fields possessing lower association fractions is due to their targeted hosts tending towards lower halo masses (and hence smaller satellite populations); meanwhile, the number of background interlopers does not vary as strongly with the target host's $M_{200}$.

We demonstrate further signs of a significant fraction of interlopers among the full sample of MATLAS dwarfs in Fig.~\ref{fig:dwarf_background}, which plots the distribution of their distance-dependent properties in gray.
The MATLAS dwarfs are assumed to lie at their presumed host galaxy's distance when calculating $M_g$ and $R_e$.
In TNG, the bound and background sample of MATLAS-like dwarfs follow similar intrinsic distributions in both $M_g$ and $R_e$.
However, the background dwarfs are fully distinct from the bound satellites when assuming they lie at a distance of $D_{\mathrm{host}}$.
Due to the different distance ranges they inhabit, the intrinsic similarity between bound and background dwarfs also results in TNG's two populations following distinct distributions in $m_g$.
W note that MATLAS dwarfs with distance estimates (black dotted lines) exhibit a significant excess at higher estimated luminosities simply because brighter dwarfs are more likely to satisfy the $\mathrm{S/N} \geq 5$ requirement for their spectra \citep{Habas2020newly} required to obtain sufficiently robust spectroscopic redshifts.

Since the MATLAS dwarf catalogue should consist of both bound satellites and interlopers, we may consider its distribution in each of these three properties as a superposition of TNG's mock-observed bound and background dwarfs at some given ratio.
For each property, we identify the fraction of satellites associated with their presumed host, $f_{\mathrm{assoc}}$, which minimises the summed CDF's Kolmogorov-Smirnov statistic (green dashed line) with respect to the observed MATLAS distribution (black dashed line outlining the grey region).
We obtained $f_{\mathrm{assoc}}$ of $0.55$, $0.62$, and $0.35$ for $M_g$, $m_g$, and $R_e$ respectively.
The optimised $R_e$ CDF does not trace the MATLAS distribution well, and its corresponding $f_{\mathrm{assoc}}$ may be unreliable.
Hence, we estimated that around $55-62\%$ of all MATLAS dwarfs are associated with their field's central ETG. This fraction is consistent with \citet{Heesters2023radial}'s result of $57\%$ (although we caution that the association fraction of the brighter MATLAS dwarfs with available distance estimates may not directly translate to the full MATLAS sample). Conversely, our estimate is based on the full MATLAS dwarf sample, and remains valid if the photometric properties of the MATLAS dwarfs are consistent with their TNG analogues.

While we could attempt to alleviate the discrepancy in dwarf population between MATLAS and the mock-observed MATLAS-like fields in TNG by introducing a dominant fraction of unassociated background galaxies in the former, it seems likely that at least one-half of the observed MATLAS dwarfs are satellite galaxies of their field's central ETG.
Overall, TNG's MATLAS-like fields demonstrate a comparable or greater degree of background contamination than MATLAS due to the lack of a visual inspection stage to remove massive interlopers with clear substructure.
The tension between MATLAS and $\Lambda$CDM expectations in TNG50 remains robust at the $>6\sigma$ level.

\subsection{The impact of simulated resolution}
\label{sec:s3_background}

The dearth of MATLAS-like dwarfs in TNG50 (reported in Section~\ref{sec:s3_populations}) may be a direct result of an insufficient simulated resolution, as well as unreliable $M_g$ and $R_e$ estimates derived from too few stellar particles.
In Fig.~\ref{fig:dwarf_resolution}, we demonstrate that the baryonic resolution in TNG50 is sufficient to resolve a majority of MATLAS-like dwarfs and yield realistic intrinsic properties.
\citet{Habas2020newly} estimates the distribution of stellar masses for 658 ($30\%$) of MATLAS dwarfs with $(g-i)$ colour estimates and clean photometry using \citet{Taylor2011galaxy}'s colour-mass relation (left panel, red line).
The dotted red line shows the $M_*$ estimates re-computed for MATLAS dwarfs with independent distance estimates.
The former consists of a fraction of background galaxies that would systematically skew the distribution towards lower $M_*$, and distances are only known for the most massive, easily observed dwarfs -- thus skewing the latter distribution towards higher $M_*$.
As expected, we find the stellar masses of TNG's mock-observed dwarfs between the two MATLAS CDFs. Finally,
$97\%$ of TNG50's bound MATLAS-like dwarfs have a stellar mass above the equivalent of 100 stellar particles.

In the right-hand panel, we show the intrinsic scaling relation of all luminous subhalos in TNG50, as well as the lower-resolution TNG100 run ($m_{\mathrm{gas}}=1.4\times10^6M_{\odot}$, $m_{\mathrm{DM}}=7.5\times10^6M_{\odot}$).
Dwarfs considered to be MATLAS-like in each run are plotted as black points.
In both runs, we identify a broadening in the $R_e$ distribution at higher $M_g$ (and lower $M_*$) due to the difficulty in estimating the half-mass radius of a subhalo containing just a few star particles.
In TNG100, this broadening begins at $M_g\sim12$ at stellar masses within an order of magnitude of most MATLAS-like dwarfs.
This is not the case for TNG50, wherein $R_e$ appears stable beyond $M_g < -10$.
We also recover a slight vertical shift in $R_e$ over the full subhalo distribution between runs, but this is likely an artefact of the larger force softening length adopted in TNG100 (185 pc) compared to TNG50 (74 pc).
Hence, we argue that TNG50 has a baryonic resolution sufficient to resolve MATLAS dwarf analogues, and the discrepancy in its simulated population with the MATLAS fields is not a direct result of unreliable $M_g$ and $\langle \mu_g \rangle$ estimates.

\section{Summary \& conclusion}
\label{sec:s4}

The MATLAS low-to-moderate density fields survey low-surface brightness dwarf galaxies around massive early-type hosts (ETGs) at distances past $10\,\mathrm{Mpc}$.
Thus, they serve as an excellent testbed for satellite galaxy abundances beyond our cosmic neighbourhood.
To check whether dwarf populations in MATLAS are consistent with expectations from $\Lambda$CDM cosmology, we mock-observed MATLAS-like fields in the high-resolution hydrodynamic simulation IllustrisTNG-50 using photometric selections faithful to the original survey.
The simulated populations of MATLAS-like dwarfs sampled in this manner follow the standard scaling relations and demonstrate photometric properties within a range consistent with the MATLAS dwarfs, especially when assuming a $30-50\%$ fraction of background interlopers unassociated with the targeted hosts.

Strikingly, the simulated fields demonstrate significantly lower abundances of MATLAS-like dwarfs than found in the survey itself.
Within isolated fields which only contain a single target ETG, MATLAS has a median of ten dwarfs with surface brightnesses of $\langle \mu_g \rangle < 26.7\,\mathrm{mag}\,\mathrm{arcsec}^{-2}$, while TNG50 only has three such dwarfs. This result constitutes the basis for the "too-many-satellites" problem in observations at the $>6\sigma$ confidence level.
We also account for the effect of host distance on the physical dimensions of the one-degree fields observed. We report that dwarf abundances within a central $175\,\mathrm{kpc}$ region around targeted ETGs demonstrate an similar tension at the $7\sigma$ confidence level.
Furthermore, when using an viewing depth of $100\,\mathrm{Mpc}$, over $50\%$ of mock-observed TNG50 dwarfs are background interlopers that are not associated with their presumed host galaxy.
On average, to match the simulated abundances of MATLAS-like dwarfs, $93\%$ of the observed MATLAS dwarfs would need to be interlopers.
This fraction is fully inconsistent with multiple estimates of the fraction of MATLAS dwarfs associated with their targeted ETG from the literature and this work, which do not drop below $50\%$.

In this work, we consider whether\ this discrepancy in dwarf populations could have arisen from systematic biases.
We find that the tension is unlikely to be a straightforward artefact of simulation resolution, since TNG50 sufficiently resolves $97\%$ of MATLAS-like dwarfs bound to their field's targeted host with at least 100 stellar particles.
The simulated dwarfs also follow the same scaling relations as found in MATLAS \citep{Habas2020newly}, ruling out any significant systematic shift in their estimated effective radii. 
The effects of extinction from dust has not been considered, but this would further reduce the abundance of MATLAS-like dwarfs recovered in TNG50.
We also do not account for the obfuscation of satellites in close projection to the central ETG, but this would similarly work to reduce TNG50's dwarf population and compound the problem.
Finally, any incompleteness in the MATLAS fields would only exacerbate the discrepancy by raising the true number of dwarfs, and the reported tension should be considered to be only a lower bound.

Our findings lie in stark contrast with \citet{Carlsten2021luminosity}, who reported the satellite luminosity functions of 12 Local Volume hosts to be fully consistent with expectations from dark matter-only cosmological simulations using a stellar-to-halo mass relation from \citet{Garrison-Kimmel2017organized} -- despite covering a similar range in host mass and satellite surface brightness.
We point out, however, that we did not assess the full luminosity function of satellites around the ATLAS$^\mathrm{3D}$ ETGs, but instead we imposed strict photometric selection cuts to match our mock-observed fields with the MATLAS observations. This resulted in a comparison between better defined sub-populations of dwarf galaxies.
Furthermore, \citet{Carlsten2021luminosity} adopted the lower-resolution TNG100 run as their fiducial hydrodynamic simulation, which struggles to resolve dwarfs with luminosities corresponding to the fainter MATLAS dwarfs.
It is also interesting to note that the range of stellar masses in which \citet{müller2024toomany} reported the luminosity function of M83 to be most discrepant with theoretical predictions ($10^7-10^8\,M_{\odot}$) coincides with the majority of MATLAS-like TNG50 dwarfs in this work.
Our results suggest that models of galaxy formation within the $\Lambda$CDM framework may still struggle to reliably reproduce dwarf abundances beyond the Local Volume. Thus, the too-many-satellites problem is extended from M83 to across 150 fields in the MATLAS catalogue.

\begin{acknowledgements}
    We thank A. Pillepich for interesting discussions and helpful inputs. K.J.K. and M.S.P. acknowledge funding via a Leibniz-Junior Research Group (project number J94/2020). M.S.P. also thanks the German Scholars Organization and Klaus Tschira Stiftung for support via a KT Boost Fund. O.M. and N.H. are grateful to the Swiss National Science Foundation for financial support under the grant number PZ00P2\_202104. We thank the anonymous referee for their constructive input which has helped us improve the manuscript.
\end{acknowledgements}

\bibliographystyle{aa}
\bibliography{paper}

\begin{thebibliography}{43}
\expandafter\ifx\csname natexlab\endcsname\relax\def\natexlab#1{#1}\fi

\bibitem[{{Bacon} {et~al.}(2010){Bacon}, {Accardo}, {Adjali}, {Anwand},
  {Bauer}, {Biswas}, {Blaizot}, {Boudon}, {Brau-Nogue}, {Brinchmann},
  {Caillier}, {Capoani}, {Carollo}, {Contini}, {Couderc}, {Daguis{\'e}},
  {Deiries}, {Delabre}, {Dreizler}, {Dubois}, {Dupieux}, {Dupuy}, {Emsellem},
  {Fechner}, {Fleischmann}, {Fran{\c{c}}ois}, {Gallou}, {Gharsa}, {Glindemann},
  {Gojak}, {Guiderdoni}, {Hansali}, {Hahn}, {Jarno}, {Kelz}, {Koehler},
  {Kosmalski}, {Laurent}, {Le Floch}, {Lilly}, {Lizon}, {Loupias}, {Manescau},
  {Monstein}, {Nicklas}, {Olaya}, {Pares}, {Pasquini}, {P{\'e}contal-Rousset},
  {Pell{\'o}}, {Petit}, {Popow}, {Reiss}, {Remillieux}, {Renault}, {Roth},
  {Rupprecht}, {Serre}, {Schaye}, {Soucail}, {Steinmetz}, {Streicher}, {Stuik},
  {Valentin}, {Vernet}, {Weilbacher}, {Wisotzki}, \& {Yerle}}]{Bacon2010muse}
{Bacon}, R., {Accardo}, M., {Adjali}, L., {et~al.} 2010, in Society of
  Photo-Optical Instrumentation Engineers (SPIE) Conference Series, Vol. 7735,
  Ground-based and Airborne Instrumentation for Astronomy III, ed. I.~S.
  {McLean}, S.~K. {Ramsay}, \& H.~{Takami}, 773508

\bibitem[{{Bertin} \& {Arnouts}(1996)}]{Bertin1996sextractor}
{Bertin}, E. \& {Arnouts}, S. 1996, \aaps, 117, 393

\bibitem[{B{\'\i}lek {et~al.}(2020)B{\'\i}lek, Duc, Cuillandre, Gwyn,
  Cappellari, Bekaert, Bonfini, Bitsakis, Paudel, Krajnovi{\'c},
  {et~al.}}]{Bilek2020census}
B{\'\i}lek, M., Duc, P.-A., Cuillandre, J.-C., {et~al.} 2020, Monthly Notices
  of the Royal Astronomical Society, 498, 2138

\bibitem[{{Bovill} \& {Ricotti}(2009)}]{Bovill2009pre-reionization}
{Bovill}, M.~S. \& {Ricotti}, M. 2009, \apj, 693, 1859

\bibitem[{{Boylan-Kolchin} {et~al.}(2011){Boylan-Kolchin}, {Bullock}, \&
  {Kaplinghat}}]{Boylan-Kolchin2011too}
{Boylan-Kolchin}, M., {Bullock}, J.~S., \& {Kaplinghat}, M. 2011, \mnras, 415,
  L40

\bibitem[{{Boylan-Kolchin} {et~al.}(2012){Boylan-Kolchin}, {Bullock}, \&
  {Kaplinghat}}]{Boylan-Kolchin2012milky}
{Boylan-Kolchin}, M., {Bullock}, J.~S., \& {Kaplinghat}, M. 2012, \mnras, 422,
  1203

\bibitem[{{Brooks} \& {Zolotov}(2014)}]{Brooks2014why}
{Brooks}, A.~M. \& {Zolotov}, A. 2014, \apj, 786, 87

\bibitem[{{Bullock} \& {Boylan-Kolchin}(2017)}]{Bullock2017small-scale}
{Bullock}, J.~S. \& {Boylan-Kolchin}, M. 2017, \araa, 55, 343

\bibitem[{{Cappellari} {et~al.}(2011){Cappellari}, {Emsellem}, {Krajnovi{\'c}},
  {McDermid}, {Scott}, {Verdoes Kleijn}, {Young}, {Alatalo}, {Bacon}, {Blitz},
  {Bois}, {Bournaud}, {Bureau}, {Davies}, {Davis}, {de Zeeuw}, {Duc},
  {Khochfar}, {Kuntschner}, {Lablanche}, {Morganti}, {Naab}, {Oosterloo},
  {Sarzi}, {Serra}, \& {Weijmans}}]{Cappellari2011atlas3d}
{Cappellari}, M., {Emsellem}, E., {Krajnovi{\'c}}, D., {et~al.} 2011, \mnras,
  413, 813

\bibitem[{{Carlsten} {et~al.}(2021){Carlsten}, {Greene}, {Peter}, {Beaton}, \&
  {Greco}}]{Carlsten2021luminosity}
{Carlsten}, S.~G., {Greene}, J.~E., {Peter}, A. H.~G., {Beaton}, R.~L., \&
  {Greco}, J.~P. 2021, \apj, 908, 109

\bibitem[{{Duc} {et~al.}(2015){Duc}, {Cuillandre}, {Karabal}, {Cappellari},
  {Alatalo}, {Blitz}, {Bournaud}, {Bureau}, {Crocker}, {Davies}, {Davis}, {de
  Zeeuw}, {Emsellem}, {Khochfar}, {Krajnovi{\'c}}, {Kuntschner}, {McDermid},
  {Michel-Dansac}, {Morganti}, {Naab}, {Oosterloo}, {Paudel}, {Sarzi}, {Scott},
  {Serra}, {Weijmans}, \& {Young}}]{Duc2015atlas3d}
{Duc}, P.-A., {Cuillandre}, J.-C., {Karabal}, E., {et~al.} 2015, \mnras, 446,
  120

\bibitem[{{Duc} {et~al.}(2014){Duc}, {Paudel}, {McDermid}, {Cuillandre},
  {Serra}, {Bournaud}, {Cappellari}, \& {Emsellem}}]{Duc2014identification}
{Duc}, P.-A., {Paudel}, S., {McDermid}, R.~M., {et~al.} 2014, \mnras, 440, 1458

\bibitem[{{Engler} {et~al.}(2021){Engler}, {Pillepich}, {Pasquali}, {Nelson},
  {Rodriguez-Gomez}, {Chua}, {Grebel}, {Springel}, {Marinacci}, {Weinberger},
  {Vogelsberger}, \& {Hernquist}}]{Engler2021abundance}
{Engler}, C., {Pillepich}, A., {Pasquali}, A., {et~al.} 2021, \mnras, 507, 4211

\bibitem[{{Ferrarese} {et~al.}(2012){Ferrarese}, {C{\^o}t{\'e}}, {Cuillandre},
  {Gwyn}, {Peng}, {MacArthur}, {Duc}, {Boselli}, {Mei}, {Erben}, {McConnachie},
  {Durrell}, {Mihos}, {Jord{\'a}n}, {Lan{\c{c}}on}, {Puzia}, {Emsellem},
  {Balogh}, {Blakeslee}, {van Waerbeke}, {Gavazzi}, {Vollmer}, {Kavelaars},
  {Woods}, {Ball}, {Boissier}, {Courteau}, {Ferriere}, {Gavazzi},
  {Hildebrandt}, {Hudelot}, {Huertas-Company}, {Liu}, {McLaughlin}, {Mellier},
  {Milkeraitis}, {Schade}, {Balkowski}, {Bournaud}, {Carlberg}, {Chapman},
  {Hoekstra}, {Peng}, {Sawicki}, {Simard}, {Taylor}, {Tully}, {van Driel},
  {Wilson}, {Burdullis}, {Mahoney}, \& {Manset}}]{Ferrarese2012next}
{Ferrarese}, L., {C{\^o}t{\'e}}, P., {Cuillandre}, J.-C., {et~al.} 2012, \apjs,
  200, 4

\bibitem[{{Garrison-Kimmel} {et~al.}(2017){Garrison-Kimmel}, {Bullock},
  {Boylan-Kolchin}, \& {Bardwell}}]{Garrison-Kimmel2017organized}
{Garrison-Kimmel}, S., {Bullock}, J.~S., {Boylan-Kolchin}, M., \& {Bardwell},
  E. 2017, \mnras, 464, 3108

\bibitem[{{Habas} {et~al.}(2020){Habas}, {Marleau}, {Duc}, {Durrell}, {Paudel},
  {Poulain}, {S{\'a}nchez-Janssen}, {Sreejith}, {Ramasawmy}, {Stemock},
  {Leach}, {Cuillandre}, {Gwyn}, {Agnello}, {B{\'\i}lek}, {Fensch},
  {M{\"u}ller}, {Peng}, \& {van der Burg}}]{Habas2020newly}
{Habas}, R., {Marleau}, F.~R., {Duc}, P.-A., {et~al.} 2020, \mnras, 491, 1901

\bibitem[{{Heesters} {et~al.}(2023){Heesters}, {M{\"u}ller}, {Marleau}, {Duc},
  {S{\'a}nchez-Janssen}, {Poulain}, {Habas}, {Lim}, \&
  {Durrell}}]{Heesters2023radial}
{Heesters}, N., {M{\"u}ller}, O., {Marleau}, F.~R., {et~al.} 2023, arXiv
  e-prints, arXiv:2305.04593

\bibitem[{{Jacobs} {et~al.}(2009){Jacobs}, {Rizzi}, {Tully}, {Shaya},
  {Makarov}, \& {Makarova}}]{Jacobs2009extragalactic}
{Jacobs}, B.~A., {Rizzi}, L., {Tully}, R.~B., {et~al.} 2009, \aj, 138, 332

\bibitem[{{Klypin} {et~al.}(1999){Klypin}, {Kravtsov}, {Valenzuela}, \&
  {Prada}}]{Klypin1999where}
{Klypin}, A., {Kravtsov}, A.~V., {Valenzuela}, O., \& {Prada}, F. 1999, \apj,
  522, 82

\bibitem[{{Kovlakas} {et~al.}(2021){Kovlakas}, {Zezas}, {Andrews}, {Basu-Zych},
  {Fragos}, {Hornschemeier}, {Kouroumpatzakis}, {Lehmer}, \&
  {Ptak}}]{Kovlakas2021heraklion}
{Kovlakas}, K., {Zezas}, A., {Andrews}, J.~J., {et~al.} 2021, \mnras, 506, 1896

\bibitem[{{Moore} {et~al.}(1999){Moore}, {Ghigna}, {Governato}, {Lake},
  {Quinn}, {Stadel}, \& {Tozzi}}]{Moore1999dark}
{Moore}, B., {Ghigna}, S., {Governato}, F., {et~al.} 1999, \apjl, 524, L19

\bibitem[{{M{\"u}ller} \& {Jerjen}(2020)}]{Müller2020abundance}
{M{\"u}ller}, O. \& {Jerjen}, H. 2020, \aap, 644, A91

\bibitem[{{M{\"u}ller} {et~al.}(2019){M{\"u}ller}, {Rejkuba}, {Pawlowski},
  {Ibata}, {Lelli}, {Hilker}, \& {Jerjen}}]{Müller2019dwarf}
{M{\"u}ller}, O., {Rejkuba}, M., {Pawlowski}, M.~S., {et~al.} 2019, \aap, 629,
  A18

\bibitem[{Müller {et~al.}(2024)Müller, Pawlowski, Revaz, Venhola, Rejkuba,
  Hilker, \& Lutz}]{müller2024toomany}
Müller, O., Pawlowski, M.~S., Revaz, Y., {et~al.} 2024
  [\eprint[arXiv]{2403.08717}]

\bibitem[{{Pawlowski} {et~al.}(2015){Pawlowski}, {Famaey}, {Merritt}, \&
  {Kroupa}}]{Pawlowski2015persistence}
{Pawlowski}, M.~S., {Famaey}, B., {Merritt}, D., \& {Kroupa}, P. 2015, \apj,
  815, 19

\bibitem[{{Pe{\~n}arrubia} {et~al.}(2010){Pe{\~n}arrubia}, {Benson}, {Walker},
  {Gilmore}, {McConnachie}, \& {Mayer}}]{Peñarrubia2010impact}
{Pe{\~n}arrubia}, J., {Benson}, A.~J., {Walker}, M.~G., {et~al.} 2010, \mnras,
  406, 1290

\bibitem[{{Peng} {et~al.}(2010){Peng}, {Ho}, {Impey}, \&
  {Rix}}]{Peng2010detailed}
{Peng}, C.~Y., {Ho}, L.~C., {Impey}, C.~D., \& {Rix}, H.-W. 2010, \aj, 139,
  2097

\bibitem[{{Pillepich} {et~al.}(2018){Pillepich}, {Springel}, {Nelson}, {Genel},
  {Naiman}, {Pakmor}, {Hernquist}, {Torrey}, {Vogelsberger}, {Weinberger}, \&
  {Marinacci}}]{Pillepich2018simulating}
{Pillepich}, A., {Springel}, V., {Nelson}, D., {et~al.} 2018, \mnras, 473, 4077

\bibitem[{{Planck Collaboration} {et~al.}(2016){Planck Collaboration}, {Ade},
  {Aghanim}, {Arnaud}, {Ashdown}, {Aumont}, {Baccigalupi}, {Banday},
  {Barreiro}, {Bartlett}, {Bartolo}, {Battaner}, {Battye}, {Benabed},
  {Beno{\^\i}t}, {Benoit-L{\'e}vy}, {Bernard}, {Bersanelli}, {Bielewicz},
  {Bock}, {Bonaldi}, {Bonavera}, {Bond}, {Borrill}, {Bouchet}, {Boulanger},
  {Bucher}, {Burigana}, {Butler}, {Calabrese}, {Cardoso}, {Catalano},
  {Challinor}, {Chamballu}, {Chary}, {Chiang}, {Chluba}, {Christensen},
  {Church}, {Clements}, {Colombi}, {Colombo}, {Combet}, {Coulais}, {Crill},
  {Curto}, {Cuttaia}, {Danese}, {Davies}, {Davis}, {de Bernardis}, {de Rosa},
  {de Zotti}, {Delabrouille}, {D{\'e}sert}, {Di Valentino}, {Dickinson},
  {Diego}, {Dolag}, {Dole}, {Donzelli}, {Dor{\'e}}, {Douspis}, {Ducout},
  {Dunkley}, {Dupac}, {Efstathiou}, {Elsner}, {En{\ss}lin}, {Eriksen},
  {Farhang}, {Fergusson}, {Finelli}, {Forni}, {Frailis}, {Fraisse},
  {Franceschi}, {Frejsel}, {Galeotta}, {Galli}, {Ganga}, {Gauthier}, {Gerbino},
  {Ghosh}, {Giard}, {Giraud-H{\'e}raud}, {Giusarma}, {Gjerl{\o}w},
  {Gonz{\'a}lez-Nuevo}, {G{\'o}rski}, {Gratton}, {Gregorio}, {Gruppuso},
  {Gudmundsson}, {Hamann}, {Hansen}, {Hanson}, {Harrison}, {Helou},
  {Henrot-Versill{\'e}}, {Hern{\'a}ndez-Monteagudo}, {Herranz}, {Hildebrandt},
  {Hivon}, {Hobson}, {Holmes}, {Hornstrup}, {Hovest}, {Huang}, {Huffenberger},
  {Hurier}, {Jaffe}, {Jaffe}, {Jones}, {Juvela}, {Keih{\"a}nen}, {Keskitalo},
  {Kisner}, {Kneissl}, {Knoche}, {Knox}, {Kunz}, {Kurki-Suonio}, {Lagache},
  {L{\"a}hteenm{\"a}ki}, {Lamarre}, {Lasenby}, {Lattanzi}, {Lawrence}, {Leahy},
  {Leonardi}, {Lesgourgues}, {Levrier}, {Lewis}, {Liguori}, {Lilje},
  {Linden-V{\o}rnle}, {L{\'o}pez-Caniego}, {Lubin}, {Mac{\'\i}as-P{\'e}rez},
  {Maggio}, {Maino}, {Mandolesi}, {Mangilli}, {Marchini}, {Maris}, {Martin},
  {Martinelli}, {Mart{\'\i}nez-Gonz{\'a}lez}, {Masi}, {Matarrese}, {McGehee},
  {Meinhold}, {Melchiorri}, {Melin}, {Mendes}, {Mennella}, {Migliaccio},
  {Millea}, {Mitra}, {Miville-Desch{\^e}nes}, {Moneti}, {Montier}, {Morgante},
  {Mortlock}, {Moss}, {Munshi}, {Murphy}, {Naselsky}, {Nati}, {Natoli},
  {Netterfield}, {N{\o}rgaard-Nielsen}, {Noviello}, {Novikov}, {Novikov},
  {Oxborrow}, {Paci}, {Pagano}, {Pajot}, {Paladini}, {Paoletti}, {Partridge},
  {Pasian}, {Patanchon}, {Pearson}, {Perdereau}, {Perotto}, {Perrotta},
  {Pettorino}, {Piacentini}, {Piat}, {Pierpaoli}, {Pietrobon}, {Plaszczynski},
  {Pointecouteau}, {Polenta}, {Popa}, {Pratt}, {Pr{\'e}zeau}, {Prunet},
  {Puget}, {Rachen}, {Reach}, {Rebolo}, {Reinecke}, {Remazeilles}, {Renault},
  {Renzi}, {Ristorcelli}, {Rocha}, {Rosset}, {Rossetti}, {Roudier},
  {Rouill{\'e} d'Orfeuil}, {Rowan-Robinson}, {Rubi{\~n}o-Mart{\'\i}n},
  {Rusholme}, {Said}, {Salvatelli}, {Salvati}, {Sandri}, {Santos},
  {Savelainen}, {Savini}, {Scott}, {Seiffert}, {Serra}, {Shellard}, {Spencer},
  {Spinelli}, {Stolyarov}, {Stompor}, {Sudiwala}, {Sunyaev}, {Sutton},
  {Suur-Uski}, {Sygnet}, {Tauber}, {Terenzi}, {Toffolatti}, {Tomasi},
  {Tristram}, {Trombetti}, {Tucci}, {Tuovinen}, {T{\"u}rler}, {Umana},
  {Valenziano}, {Valiviita}, {Van Tent}, {Vielva}, {Villa}, {Wade}, {Wandelt},
  {Wehus}, {White}, {White}, {Wilkinson}, {Yvon}, {Zacchei}, \&
  {Zonca}}]{PlanckCollaboration2016}
{Planck Collaboration}, {Ade}, P.~A.~R., {Aghanim}, N., {et~al.} 2016, \aap,
  594, A13

\bibitem[{{Poulain} {et~al.}(2021){Poulain}, {Marleau}, {Habas}, {Duc},
  {S{\'a}nchez-Janssen}, {Durrell}, {Paudel}, {Ahad}, {Chougule}, {M{\"u}ller},
  {Lim}, {B{\'\i}lek}, \& {Fensch}}]{Poulain2021structure}
{Poulain}, M., {Marleau}, F.~R., {Habas}, R., {et~al.} 2021, \mnras, 506, 5494

\bibitem[{{Price} {et~al.}(2022){Price}, {{\"U}bler}, {F{\"o}rster Schreiber},
  {de Zeeuw}, {Burkert}, {Genzel}, {Tacconi}, {Davies}, \&
  {Price}}]{Price2022kinematics}
{Price}, S.~H., {{\"U}bler}, H., {F{\"o}rster Schreiber}, N.~M., {et~al.} 2022,
  \aap, 665, A159

\bibitem[{{Radburn-Smith} {et~al.}(2011){Radburn-Smith}, {de Jong}, {Seth},
  {Bailin}, {Bell}, {Brown}, {Bullock}, {Courteau}, {Dalcanton}, {Ferguson},
  {Goudfrooij}, {Holfeltz}, {Holwerda}, {Purcell}, {Sick}, {Streich}, {Vlajic},
  \& {Zucker}}]{Radburn-Smith2011ghosts}
{Radburn-Smith}, D.~J., {de Jong}, R.~S., {Seth}, A.~C., {et~al.} 2011, \apjs,
  195, 18

\bibitem[{{Revaz} \& {Jablonka}(2018)}]{Revaz2018pushing}
{Revaz}, Y. \& {Jablonka}, P. 2018, \aap, 616, A96

\bibitem[{{Sales} {et~al.}(2022){Sales}, {Wetzel}, \&
  {Fattahi}}]{Sales2022baryonic}
{Sales}, L.~V., {Wetzel}, A., \& {Fattahi}, A. 2022, Nature Astronomy, 6, 897

\bibitem[{{Samuel} {et~al.}(2020){Samuel}, {Wetzel}, {Tollerud},
  {Garrison-Kimmel}, {Loebman}, {El-Badry}, {Hopkins}, {Boylan-Kolchin},
  {Faucher-Gigu{\`e}re}, {Bullock}, {Benincasa}, \&
  {Bailin}}]{Samuel2020profile}
{Samuel}, J., {Wetzel}, A., {Tollerud}, E., {et~al.} 2020, \mnras, 491, 1471

\bibitem[{{Sawala} {et~al.}(2016){Sawala}, {Frenk}, {Fattahi}, {Navarro},
  {Bower}, {Crain}, {Dalla Vecchia}, {Furlong}, {Helly}, {Jenkins}, {Oman},
  {Schaller}, {Schaye}, {Theuns}, {Trayford}, \& {White}}]{Sawala2016apostle}
{Sawala}, T., {Frenk}, C.~S., {Fattahi}, A., {et~al.} 2016, \mnras, 457, 1931

\bibitem[{{Schaye} {et~al.}(2015){Schaye}, {Crain}, {Bower}, {Furlong},
  {Schaller}, {Theuns}, {Dalla Vecchia}, {Frenk}, {McCarthy}, {Helly},
  {Jenkins}, {Rosas-Guevara}, {White}, {Baes}, {Booth}, {Camps}, {Navarro},
  {Qu}, {Rahmati}, {Sawala}, {Thomas}, \& {Trayford}}]{Schaye2015eagle}
{Schaye}, J., {Crain}, R.~A., {Bower}, R.~G., {et~al.} 2015, \mnras, 446, 521

\bibitem[{{Smercina} {et~al.}(2018){Smercina}, {Bell}, {Price}, {D'Souza},
  {Slater}, {Bailin}, {Monachesi}, \& {Nidever}}]{Smercina2018lonely}
{Smercina}, A., {Bell}, E.~F., {Price}, P.~A., {et~al.} 2018, \apj, 863, 152

\bibitem[{{Spergel} {et~al.}(2007){Spergel}, {Bean}, {Dor{\'e}}, {Nolta},
  {Bennett}, {Dunkley}, {Hinshaw}, {Jarosik}, {Komatsu}, {Page}, {Peiris},
  {Verde}, {Halpern}, {Hill}, {Kogut}, {Limon}, {Meyer}, {Odegard}, {Tucker},
  {Weiland}, {Wollack}, \& {Wright}}]{Spergel2007three-year}
{Spergel}, D.~N., {Bean}, R., {Dor{\'e}}, O., {et~al.} 2007, \apjs, 170, 377

\bibitem[{{Springel} {et~al.}(2001){Springel}, {White}, {Tormen}, \&
  {Kauffmann}}]{Springel2001populating}
{Springel}, V., {White}, S. D.~M., {Tormen}, G., \& {Kauffmann}, G. 2001,
  \mnras, 328, 726

\bibitem[{{Taylor} {et~al.}(2011){Taylor}, {Hopkins}, {Baldry}, {Brown},
  {Driver}, {Kelvin}, {Hill}, {Robotham}, {Bland-Hawthorn}, {Jones}, {Sharp},
  {Thomas}, {Liske}, {Loveday}, {Norberg}, {Peacock}, {Bamford}, {Brough},
  {Colless}, {Cameron}, {Conselice}, {Croom}, {Frenk}, {Gunawardhana},
  {Kuijken}, {Nichol}, {Parkinson}, {Phillipps}, {Pimbblet}, {Popescu},
  {Prescott}, {Sutherland}, {Tuffs}, {van Kampen}, \&
  {Wijesinghe}}]{Taylor2011galaxy}
{Taylor}, E.~N., {Hopkins}, A.~M., {Baldry}, I.~K., {et~al.} 2011, \mnras, 418,
  1587

\bibitem[{{Tempel} {et~al.}(2014){Tempel}, {Stoica}, {Mart{\'\i}nez},
  {Liivam{\"a}gi}, {Castellan}, \& {Saar}}]{Tempel2014detecting}
{Tempel}, E., {Stoica}, R.~S., {Mart{\'\i}nez}, V.~J., {et~al.} 2014, \mnras,
  438, 3465

\bibitem[{{Zana} {et~al.}(2022){Zana}, {Lupi}, {Bonetti}, {Dotti},
  {Rosas-Guevara}, {Izquierdo-Villalba}, {Bonoli}, {Hernquist}, \&
  {Nelson}}]{Zana2022morphological}
{Zana}, T., {Lupi}, A., {Bonetti}, M., {et~al.} 2022, \mnras, 515, 1524

\end{thebibliography}

\begin{appendix}

\section{Impact of host isolation}
\label{app:isolation}

\begin{figure}
    \centering
    \includegraphics[width=\columnwidth]{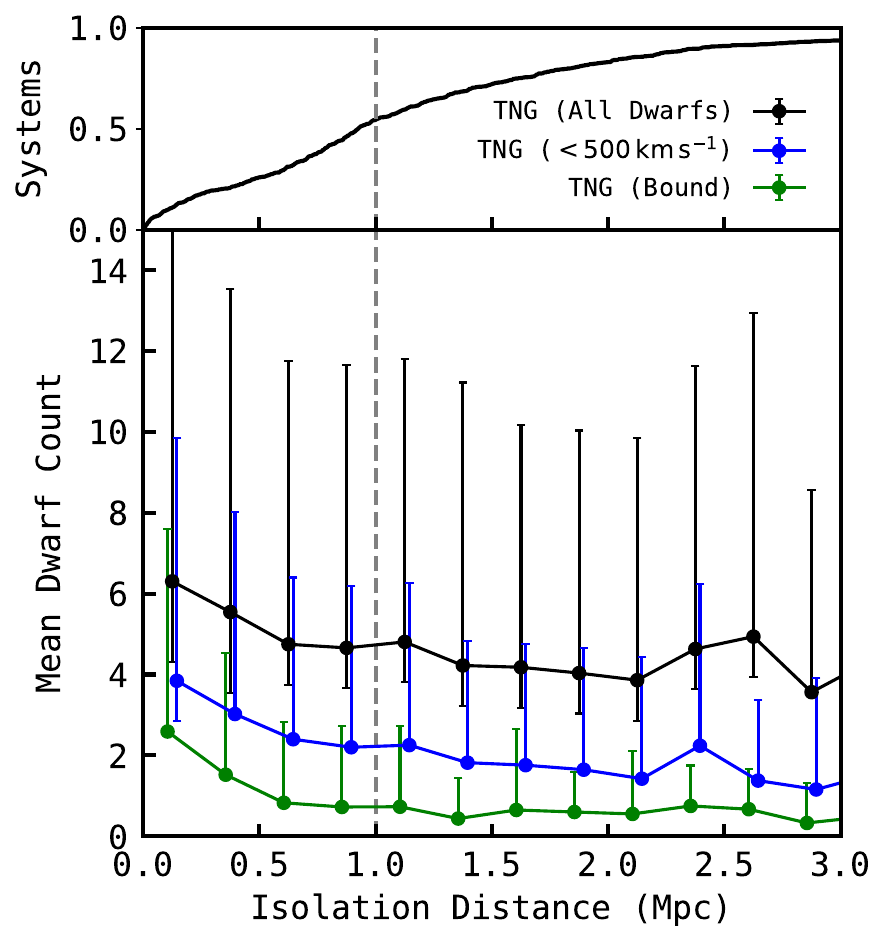}
    \caption{
    Mean dwarf populations of isolated MATLAS-like TNG50 fields, binned by the 3D distance between target hosts and their closest bright companion with a magnitude of $M_K + 1$.
    Black, blue, and green points correspond to samples of all MATLAS-like dwarfs, those with a radial velocity within $500\,\mathrm{km}\,\mathrm{s}^{-1}$ of their presumed host, and dwarfs considered to be gravitationally bound to their host's halo in \texttt{Subfind} -- error bars in their respective colours denote their $1\sigma$ spread.
    The upper panel shows the cumulative distribution of mock-observed MATLAS-like field realisations.
    This work's adopted isolation criterion of $1\,\mathrm{Mpc}$ is indicated by the grey dashed line.
    }
    \label{fig:isol_distance}
\end{figure}

We verified whether the results obtained in this work is sensitive to our adopted isolation criterion of $1\,\mathrm{Mpc}$ imposed on the host galaxies mock-observed in TNG50.
In Fig.~\ref{fig:isol_distance}, we plot the mean dwarf counts in MATLAS-like fields as a function of their hosts' distance to their closest companion.
Until around $600\,\mathrm{kpc}$, where the satellites of the companion host begin to encroach on the target's dwarf population, the number of MATLAS-like dwarfs within the mock-observed fields (especially those physically associated to their presumed host galaxy) remains relatively stable.
We conclude that our specific choice of $1\,\mathrm{Mpc}$ in our isolation criterion is unlikely to significantly contribute to the observed discrepancy in dwarf populations. Imposing a stricter degree of isolation would only serve to artificially reduce the significance of the discrepancy due to the smaller sample of MATLAS fields.

\end{appendix}

\end{document}